\documentclass[%
reprint,
superscriptaddress,
showpacs,
 amsmath,amssymb,
 aps,
prb,
floatfix,
]{revtex4-1}

\usepackage{graphicx}
\usepackage{dcolumn}
\usepackage{bm}
\usepackage{xfrac}
\usepackage{color,soul}
\usepackage{multirow}
\usepackage{booktabs}
\usepackage{float}
\usepackage{makecell}
\usepackage{siunitx}
\usepackage{threeparttable}
\usepackage{enumerate}

\usepackage[final]{changes}

\usepackage{python}

\newcommand{\mbfr}{\mathbf{r}}
\newcommand{\dd}{\mathrm{d}}

\newcommand{\bra}[1]{\langle #1 |}
\newcommand{\ket}[1]{| #1 \rangle}
\newcommand{\braket}[2]{\langle #1 | #2 \rangle}

\newcommand{\CASTEP}{CASTEP}
\newcommand{\mbj}{TB}
\newcommand{\etal}{\emph{et al}}
\newcommand{\myred}{\color{red}}
\definechangesauthor[color=red]{ABP}

\begin{document}


\title{Ultrasoft pseudopotentials with kinetic energy density support: implementing the Tran-Blaha potential}

\author{Albert P. Bart\'ok}
\affiliation{%
Scientific Computing Department\\
Science and Technology Facilities Council, Rutherford Appleton Laboratory, Didcot, OX11 0QX, United Kingdom}
 \email{apbartok@gmail.com}
\author{Jonathan R. Yates}%
\affiliation{%
Department of Materials\\
University of Oxford, Oxford OX1 3PH, United Kingdom}%

\date{\today}

\begin{abstract}
We extend the Vanderbilt ultrasoft pseudopotential scheme by adding kinetic energy density terms, in order to use meta-GGA exchange potentials, such as the Becke-Johnson or Tran-Blaha potentials, in the planewave-pseudopotential implementation of Density Functional Theory. Having implemented kinetic energy augmentation and non-linear core correction terms in the \CASTEP{} density functional package, we evaluate the validity of our approach by comparing the calculated electronic structure of isolated atoms and semiconductor crystals to all-electron benchmark calculations. Based on our results, we provide recommendations for the practical use of the Tran-Blaha exchange in planewave-pseudopotential codes.
\end{abstract}

\pacs{71.15.Mb,71.20.Mq,71.20.Nr,71.15.Ap,71.15.Dx}
\maketitle


\section{\label{sec:intro}Introduction}
Density functional theory (DFT) calculations have become a standard tool in atomistic modelling, resulting from a good balance between computational cost and accuracy.
Due to the approximations in practical applications, DFT is often labelled less accurate than high level quantum chemistry methods. On the other hand, DFT is generally regarded as a good model of atomic interactions and often used to gain insight of structures and dynamics on the microscopic level. Properties derived from the the electronic ground state are also well described by DFT, as demonstrated by the successful prediction of vibrational properties\cite{Baroni:2001tn} and Nuclear Magnetic Resonance (NMR) parameters from first principles.\cite{Bonhomme:2012cz} Most commonly used in practical applications, the Kohn-Sham (KS) equations provide a way to map the all-electron problem to a set of non-interacting one-electron Schr\"odinger equations,\cite{Kohn:1965uia} where the kinetic energy is well defined albeit not exact for the many-body interacting system. The difference can be formulated in terms of exchange and correlation functionals, which are approximated in practical calculations.

One of the most often used approaches, the Generalized Gradient Approximation (GGA) owes its success over the Local Density Approximation (LDA) to its dependence on the local gradient of the electronic density in addition to the value of the density, allowing more accurate description of variations in the electron-electron interactions. However, missing description of self-interaction\cite{Bao:2018ei} and the derivative discontinuity with respect to particle numbers\cite{Armiento:2013ed} result in an inferior description of the band structure \added{by underestimating the band gaps} of solids compared to higher level of theory, such as \replaced{GW, which approximates the self-energy using the single-particle Green's function (G) and the screened Coulomb interaction (W)}{GW}\cite{doi:10.1002/wcms.1344} \deleted{or TDDFT by underestimating the band gaps}.  Some GGA parameterizations such as PBE are known to overestimate bond lengths while underestimating bonding energy. Several approaches have been proposed to improve these shortcomings. Hybrid functionals use exact exchange obtained from the single determinant exchange using the Kohn-Sham orbitals, but in practice the amount of exact exchange is fitted to reproduce a select group of properties, and as a result, they lack generality. Meta-GGA (mGGA) functionals, on the other hand, follow the GGA idea, but including higher order derivatives of the electronic density or the kinetic energy density (KED), thereby introducing more non-local effects, in the fashion of a Taylor-expansion. It has been shown that SCAN\cite{Sun:2015ef}, a mGGA functional improves considerably the description of atomic interactions as well as the electronic structure, if the generalized Kohn-Sham scheme is used to determine the electronic ground state\cite{Yang:2016ef}. Separate developments related to mGGA, for example, the Becke-Johnson potential\cite{Becke:2006ky} (BJ) and its modification proposed by Tran and Blaha\cite{Tran:2009kk} (TB), aim to fix the problem of underestimation of band-gaps.

Implementing mGGA functionals is straightforward in all-electron DFT codes, and many software packages already allow such calculations\cite{elk,wien2k,g16}, but in the planewave-pseudopotential framework additional considerations are needed for KED terms. Sun \etal{} have described the changes required in the Projector Augmented Wave method to enable self-consistent mGGA calculations\cite{Sun:2011ka}. Yao and Kanai discussed implementation details of norm-conserving pseudopotentials with the mGGA functionals\cite{Yao:2017gn}, which are, however missing KED augmentation terms and non-linear core corrections.

The TB potential has been used in plane-wave pseudopotential codes in previous studies\cite{Waroquiers:2013je}, albeit with inconsistent pseudopotentials, generated with LDA or GGA functionals\cite{Germaneau:2013hl,Sato:2015dg}. Germaneau \etal{} found that the accuracy of the resulting band gaps depends on the pseudopotentials, and recommended GGA pseudopotentials.\cite{Germaneau:2013hl}

In this work, we explore how the potential-only mGGA methodology can be implemented in the plane-wave, ultrasoft pseudopotential framework of DFT. The use of this class of effective potentials is somewhat limited, due to the fact that there is no corresponding functional defined, therefore neither the total energy, nor structural or thermochemical information are available. However, there has been a considerable interest in the technique, and our work aims to increase the accuracy of planewave pseudopotential calculations within the potential-only framework. We describe a method to generate consistent pseudopotentials, and discuss how the all-electron KED can be reconstructed and represented in a basis of plane-waves. We benchmarked our approach against the all-electron implementation and we remark on the practical limitations of the Tran-Blaha (TB) potential when using pseudopotentials.

\section{Implementation}

The \mbj{} potential\cite{Tran:2009kk} is defined as
\begin{equation}
\label{eq:MBJ}
v^{\mathrm{\mbj{}}}_{x,\sigma}(\mbfr{}) = c v^{\mathrm{BR}}_{x,\sigma}(\mbfr{}) +
(3c-2)\frac{1}{\pi} \sqrt{\frac{5}{12}} \sqrt{\frac{2 t_\sigma(\mbfr{})}{\rho_\sigma(\mbfr{})}}
\textrm{,}
\end{equation}
where $\rho_\sigma(\mbfr{})$ is the electronic density, $t_\sigma(\mbfr{})$ the kinetic energy density, $v^\mathrm{BR}_{x,\sigma} (\mbfr{})$ is the Becke-Roussel potential, and $c$ is a constant.
The electronic density is \replaced{obtained from the Kohn-Sham eigen-states $\psi_{i,\sigma}$ corresponding to spin state $\sigma$, and the occupation numbers $f_{i,\sigma}$ as}{given by}
\begin{equation}
\rho_\sigma(\mbfr{}) = \sum_i^{N_\sigma} f_{i,\sigma} |\psi_{i,\sigma}(\mbfr{})|^2
\textrm{.}
\end{equation}
 
\replaced{The Kohn-Sham}{and the} kinetic energy density is
\begin{equation}
\label{eq:ked}
t_\sigma(\mbfr{}) = \frac{1}{2} \sum_i^{N_\sigma} f_i \nabla \psi_{i,\sigma}^*(\mbfr{}) \nabla \psi_{i,\sigma}(\mbfr{})
\end{equation}
or alternatively,
\begin{equation}
t_\sigma(\mbfr{}) =  \frac{1}{4} \nabla^2 \rho_\sigma(\mbfr{}) - \frac{1}{2}\sum_i^{N_\sigma} f_i \operatorname{Re}\Bigl( \nabla^2 \psi_{i,\sigma}^*(\mbfr{}) \psi_{i,\sigma}(\mbfr{})  \Bigr)
\textrm{.}
\end{equation}
The Becke-Roussel part\cite{Becke:1989vo} of the the \mbj{} potential is given by the expression
\begin{equation}
v^\mathrm{BR}_{x,\sigma} (\mbfr{})= -\frac{1}{b_\sigma(\mbfr{})} \Biggl( 1 - e^{-x_\sigma(\mbfr{})} 
- \frac{1}{2} x_\sigma(\mbfr{}) e^{-x_\sigma(\mbfr{})} \Biggr)
\end{equation}
where $b$ is defined as
\begin{equation}
b_\sigma(\mbfr{}) = \Biggl[ \frac{x^3_\sigma (\mbfr{}) e^{-x_\sigma(\mbfr{})}}{8 \pi \rho_\sigma(\mbfr{})} \Biggr]^{\frac{1}{3}}
\textrm{.}
\end{equation}
To obtain $x_\sigma$, the non-linear equation
\begin{equation}
\frac{x_\sigma e^{\sfrac{-2x_\sigma}{3}}}{x_\sigma-2} = \frac{2}{3} \pi^{\sfrac{2}{3}} \frac{\rho_\sigma^{\sfrac{5}{3}}} {Q_\sigma}
\end{equation}
need to be solved, with the definitions
\begin{equation}
Q_\sigma = \frac{1}{6} (\nabla^2 \rho_\sigma - 2 \gamma D_\sigma)
\end{equation}
and
\begin{equation}
D_\sigma = t_\sigma - \frac{1}{4} \frac{ (\nabla \rho_\sigma)^2 }{\rho_\sigma}
\textrm{,}
\end{equation}
where we use $\gamma=0.8$\cite{Becke:1989vo}.
With $c=1$, we recover the original Becke-Johnson expression for the exchange functional. Tran and Blaha introduced a dependence of $c$ on the electronic density as
\begin{equation}
c = \alpha + \beta \Biggl( \frac{1}{V_\mathrm{cell}} \int_\mathrm{cell} \dd{} \mbfr{} \frac{|\nabla \rho(\mbfr{})|}{\rho(\mbfr{})} \Biggr)^\frac{1}{2}
\label{eq:TB_c}
\end{equation}
where they fitted the parameters $\alpha$ and $\beta$ to reproduce the band gaps of a wide range of solids.

We implemented the \mbj{} potential in the planewave DFT program, \CASTEP{}\cite{Clark:2005vp}. As the electronic density is represented on a fine Fourier grid, the gradient and the Laplacian of the electronic density are easily available. To compute the KED of the all-electron wave function, we adapt the Vanderbilt ultrasoft pseudopotential method\cite{Vanderbilt:1990uj,Laasonen:1993wg}.

\subsection{Kinetic energy density augmentation}

In order to accurately evaluate the mGGA exchange-correlation potential, the KED of all electrons is required, whereas in plane-wave codes only the valance electrons are treated explicitly, with the region near the nucleus pseudized. In the case of ultrasoft pseudopotentials, the valence electron KED can be reconstructed from the smooth pseudo-wave functions using projector and augmentation functions. We discuss the necessary kinetic energy density augmentation in terms of the Projector Augmented Wave (PAW) approach\cite{Blochl:1994uk}, noting that to generate ultrasoft pseudopotentials, the augmentation functions are pseudized close to the nucleus.

The \added{full-potential KS orbital, sometimes referred to as the} all-electron wave function\cite{Kresse:1999wc}, is reconstructed from the soft wave function using PAW as
\begin{equation}
\ket{\psi_n}= \ket{\tilde{\psi}_n} + \sum_i (\ket{\phi_i}-\ket{\tilde{\phi}_i}) \braket{\beta_i}{\tilde{\psi}_n} 
\textrm{,}
\end{equation}
where $\ket{\tilde{\psi}_n}$ are the nodeless pseudo wave functions, $\ket{\phi_i}$ are the all-electron partial waves, $\ket{\tilde{\phi}_i}$ are the pseudo partial waves and $\ket{\beta_i}$ are the projector functions\added{, as introduced by Vanderbilt\cite{Vanderbilt:1990uj} and Laasonen \etal\cite{Laasonen:1993wg}}. For clarity, we dropped the spin index $\sigma$, noting that it can be reintroduced later.

Following the derivation in \cite{Laasonen:1993wg,Sun:2011ka} the all-electron KED is reconstructed by
\begin{widetext}
\begin{equation}
t = \sum_n f_n \Biggl[ \braket{\tilde{\psi}_n}{\nabla} \braket{\nabla}{\tilde{\psi}_n} + 
\sum_{ij}
\braket{\tilde{\psi}_n}{\beta_i}\braket{\beta_j}{\tilde{\psi}_n}
\Bigl( \braket{\phi_i}{\nabla}\braket{\nabla}{\phi_j}-
\braket{\tilde{\phi}_i}{\nabla}\braket{\nabla}{\tilde{\phi}_j} \Bigr) \Biggr]
\end{equation}
\end{widetext}
\added{where $\ket{\nabla}\bra{\nabla}$ is the KED operator and} we define the KED augmentation term as
\begin{equation}
    T_{ij}(\mbfr{}) \equiv \braket{\phi_i}{\nabla}\braket{\nabla}{\phi_j}-
\braket{\tilde{\phi}_i}{\nabla}\braket{\nabla}{\tilde{\phi}_j}
\textrm{.}
\end{equation}
The partial waves $\phi$ and $\tilde{\phi}$ are expressed as the product of radial and spherical harmonics functions:
\begin{equation}
\phi_i(\mbfr{}) = \phi_i(r) Y_{l_i m_i}(\hat{\mbfr{}})
\end{equation}
The product $\phi^*_i(\mbfr{})\phi_j(\mbfr{})$ can be written as
\begin{multline}
\phi_i^*(\mbfr{}) \phi_j(\mbfr{}) = \phi^*_i(r) \phi_j(r) Y^*_{l_i m_i}(\hat{\mbfr{}}) Y_{l_j m_j}(\hat{\mbfr{}}) = \\
\sum_{LM}c_{ijLM}(r) Y_{LM}(\hat{\mbfr{}})
\textrm{,}
\end{multline}
thanks to the Clebsch-Gordan (CG) expansion of spherical harmonics.
Rearranging the Laplacian of this product
\begin{multline}
\nabla \phi_i^*(\mbfr{}) \cdot \nabla \phi_j(\mbfr{}) = \frac{1}{2} \Biggl[
\nabla^2 \Bigl(\phi_i^*(\mbfr{}) \phi_j(\mbfr{}) \Bigr) - \\
\nabla^2 \phi_i^*(\mbfr{}) \phi_j(\mbfr{}) - \phi_i^*(\mbfr{}) \nabla^2 \phi_j(\mbfr{}) \Biggr]
\end{multline}
we obtain the expression for the KED augmentation in the form of
\begin{equation}
T_{ij}(\mbfr{}) = \sum_{LM} t_{ijLM}(r) Y_{LM}(\hat{\mbfr{}})
\end{equation}
exploiting the fact that the Laplacian operator in terms of spherical coordinates is
\begin{equation}
\nabla^2 = \frac{1}{r^2} \Biggl[ \frac{\partial}{\partial r} \Bigl( r^2 \frac{\partial}{\partial r} \Bigr) - \hat{l}^2 \Biggr]
\end{equation}
and by further application of the CG expansion.

To represent the augmented KED on a Fourier grid, it would need to be very dense to be able to account for the rapid variations of the KED near the nucleus, making calculations impractical. Instead, we pseudize the $t_{ijLM}(r)$ functions in a similar way as the charge density augmentation functions are pseudized in Ref. \onlinecite{Laasonen:1993wg}, except we do not enforce conservation of moments, just optimal smoothness and the continuation conditions at the inner cutoff radius $r_\textrm{in}$.

The pseudized KED augmentation terms $\tilde{t}$ outside $r_\textrm{in}$ match exactly $t$, and inside $r_\textrm{in}$ are expanded as polynomials
\begin{equation}
\label{eq:pseudo_t}
\tilde{t}_{ijLM}(r) = \sum_{k=0}^{n} c_k r^{L+2k} \textrm{,}
\end{equation}
where $n$ is chosen such that the resulting polynomial is smooth and joining conditions regarding the \added{$\alpha$-th} derivatives at $r_\textrm{in}$ can be fulfilled:
\begin{equation}
\label{eq:t_deriv}
    \tilde{t}^{(\alpha)} (r_\textrm{in}) = t^{(\alpha)} (r_\textrm{in}), \quad \alpha=0,1,\ldots
\end{equation}
\added{In our implementation in \CASTEP{}, we preserve up to third order derivatives at $r_\textrm{in}$.} The coefficients are determined by the requirement that the Fourier coefficients of $\tilde{t}$ above a plane wave cutoff $G_\textrm{cut}$ regarding the dense Fourier grid in the plane wave calculation are as small as possible, minimizing
\begin{equation}
\label{eq:t_coeff}
    I = \int_{G_\textrm{cut}}^\infty \! \dd G \, G^2 \, \tilde{t}^2(G) \textrm{.}
\end{equation}
\added{As the pseudized KED augmentation functions, unlike their electron density counterparts, do not need to conserve the moments of the KED, they tend to be smoother. We have found that the $G_\textrm{cut}$ values applied to the electron density augmentation functions remain adequate choices for $\tilde{t}(r)$.}
\begin{figure}[ht]
    \includegraphics[width=\linewidth]{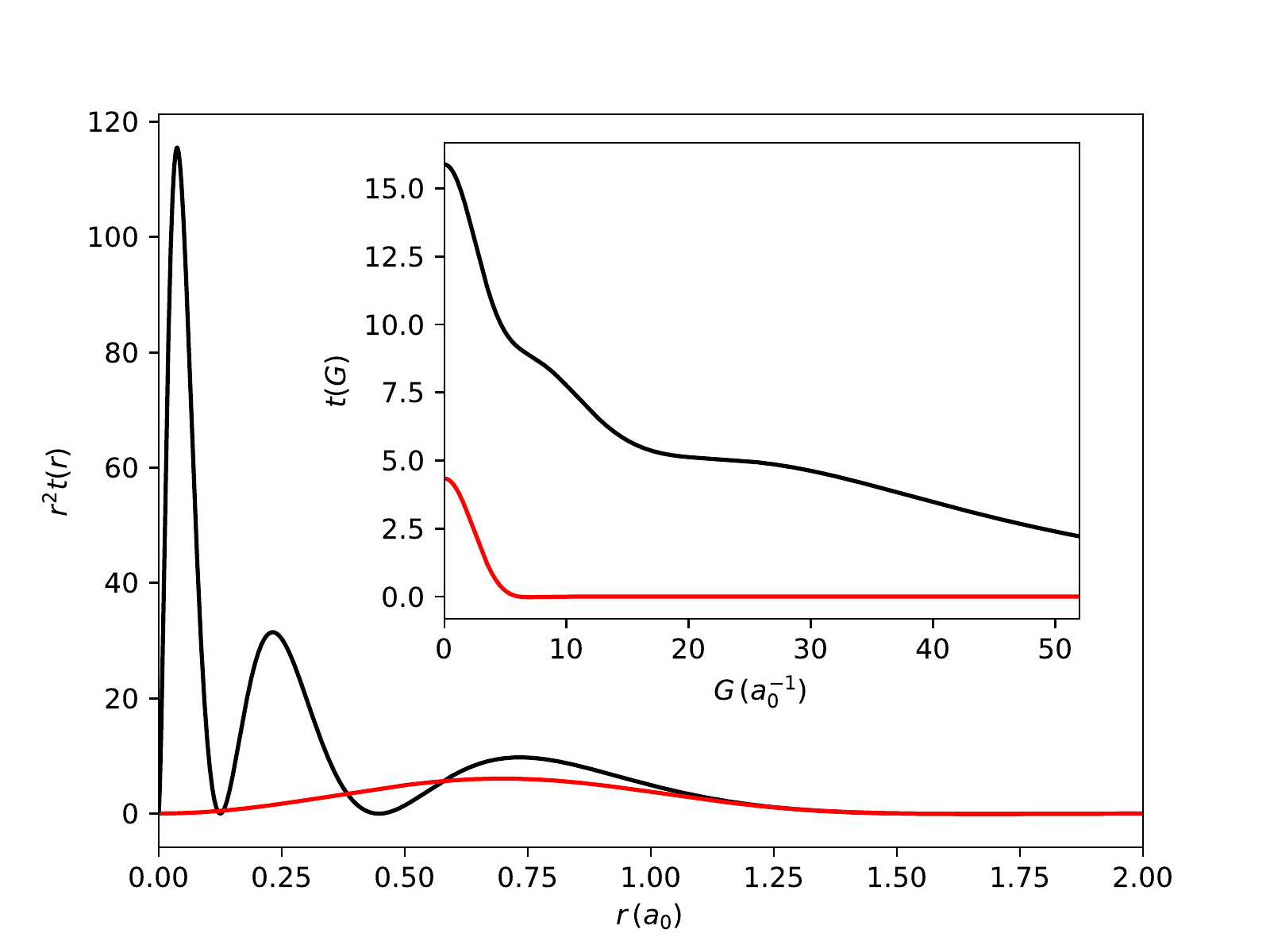}
    \includegraphics[width=\linewidth]{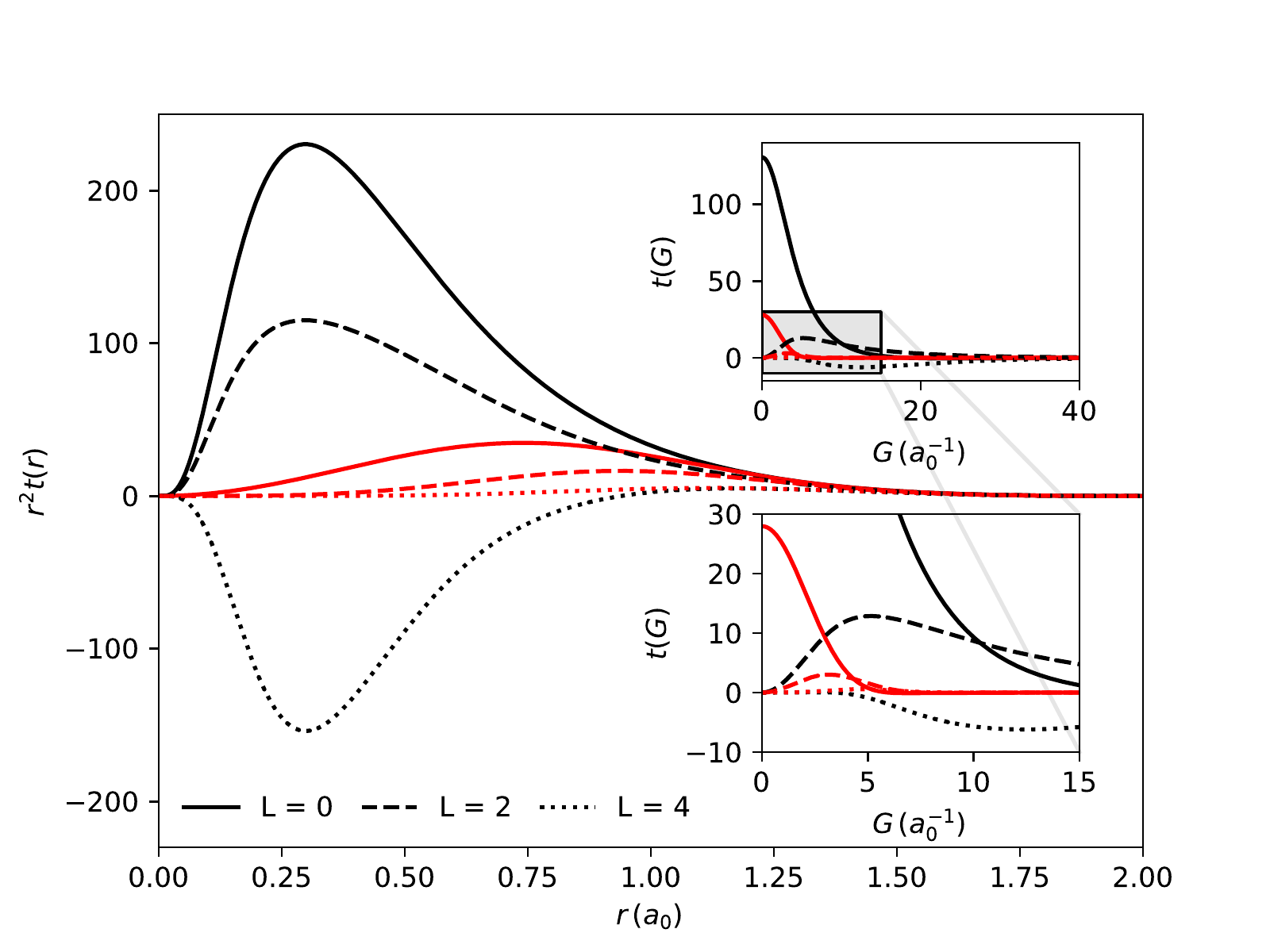}
    \caption{The radial part of the KED augmentation function belonging to the 4s (top panel) and 3d (bottom panel) orbitals of Zn, for each angular momentum channel. Black and red lines correspond to the original and the pseudized functions, respectively. The main figures show the functions in real space, whereas the insets show them in Fourier representation.}
    \label{fig:t_aug}
\end{figure}

Figure~\ref{fig:t_aug} shows an example of KED augmentation functions of the Zn atom where the pseudization radius was chosen to be $r_\textrm{in} = 1.406 \, a_0$. It is apparent that the KED augmentation functions would need a very fine grid spacing if they were to be represented accurately on a Fourier grid, which would make a calculation require impractically large memory and computational time. However, their pseudized counterparts provide a much more favourable reciprocal space convergence, with computational requirements comparable to those of GGA functionals.

Another crucial component to generate successful pseudopotentials is including a non-linear core correction (NLCC) for the KED. To accomplish this, we calculate the contribution of the core orbitals to the KED, which again need to be pseudized within a radius for practical calculations. The rapidly varying part of the function, close to the nucleus, is replaced by a smooth curve, using the same procedure described earlier in this section, by equations~\ref{eq:pseudo_t}, \ref{eq:t_deriv} and \ref{eq:t_coeff} and setting $L=0$, as the contribution from core electrons is spherically symmetric. We use the NLCC for the kinetic energy density in an similar fashion to NLCC of the charge density.\cite{Louie:1982kr}

\section{Results}

We first performed benchmark calculations on our KED pseudopotential generation scheme against all-electron calculations to establish its validity. Our tests included isolated atoms and various properties of condensed systems. In all our calculations with the Tran-Blaha or Becke-Johnson exchange potential, we used the correlation part of  LDA\cite{Perdew:1981wm}. \added{The pseudopotentials were generated using the on-the-fly scheme implemented in the \CASTEP{} code with the modifications described in the previous section. We used the pseudopotential parameters defined in the}  \verb+C17+ \added{library in \CASTEP{}.}

\subsection{Calculations on isolated atoms}

We studied the energy levels of isolated atoms and compared the self-consistent exchange potentials of the all-electron and pseudo calculations. We solved the KS equations on a logarithmic radial grid using the atomic solvers built in the \CASTEP{} code, with the relativistic effects treated by the technique suggested by Koelling and Harmon\cite{Koelling:2001jn}. The Tran-Blaha exchange potential is ill-defined for systems containing large voids, due to the construction of $c$ as an integral of the cell volume: the choice of volume is arbitrary. This is indeed the case of isolated atoms, hence we opted for the original Becke-Johnson potential\cite{Becke:2006ky} in this test, expressed as $c=1$ in equation~\ref{eq:MBJ}. We also compared the exchange-correlation potential of the pseudo atom to the all-electron solution to establish the validity of our pseudopotential scheme.

\begin{table}[h]
\begin{tabular}{ll|rrr}
\toprule
   &    &  AE (eV) &  PS (eV) & $\Delta$ (eV) \\
\hline
\hline
\midrule
\multirow{2}{*}{Be} & 1s & -110.143 & -110.119 &    -0.024 \\
   & 2s &   -5.915 &   -5.915 &    0.000 \\
\cline{1-5}
\multirow{2}{*}{Ne} & 2s &  -38.209 &  -38.208 &    -0.001 \\
   & 2p &  -14.914 &  -14.912 &    -0.002 \\
\cline{1-5}
\multirow{3}{*}{Mg} & 2s &  -82.396 &  -82.386 &    -0.010 \\
   & 2p &  -48.985 &  -48.974 &    -0.011 \\
   & 3s &   -4.997 &   -4.997 &    0.000 \\
\cline{1-5}
\multirow{2}{*}{Ar} & 3s &  -25.109 &  -25.108 &    -0.001 \\
   & 3p &  -11.353 &  -11.353 &    0.000 \\
\cline{1-5}
\multirow{3}{*}{Ca} & 3s &  -47.820 &  -47.818 &    -0.002 \\
   & 3p &  -29.168 &  -29.167 &    -0.001 \\
   & 4s &   -3.990 &   -3.990 &    0.000 \\
\cline{1-5}
\multirow{2}{*}{Zn} & 3d &  -11.703 &  -11.701 &    -0.002 \\
   & 4s &   -6.084 &   -6.084 &     0.000 \\
\cline{1-5}
\multirow{2}{*}{Kr} & 4s &  -23.123 &  -23.123 &    0.000 \\
   & 4p &  -10.248 &  -10.248 &    0.000 \\
\cline{1-5}
\multirow{2}{*}{Cd} & 4d &  -13.649 &  -13.648 &    -0.001 \\
   & 5s &   -5.663 &   -5.664 &     0.000 \\
\bottomrule
\end{tabular}
    \caption{Comparison of energy levels of valence orbitals of neutral and isolated atoms, calculated considering all electrons (AE) and the ultrasoft pseudopotential scheme (PS), using the Becke-Johnson exchange potential.}
    \label{tab:singe_atom_energy}
\end{table}

\begin{figure*}[htb]
\begin{tabular}{cc}
\includegraphics[width=0.8\columnwidth]{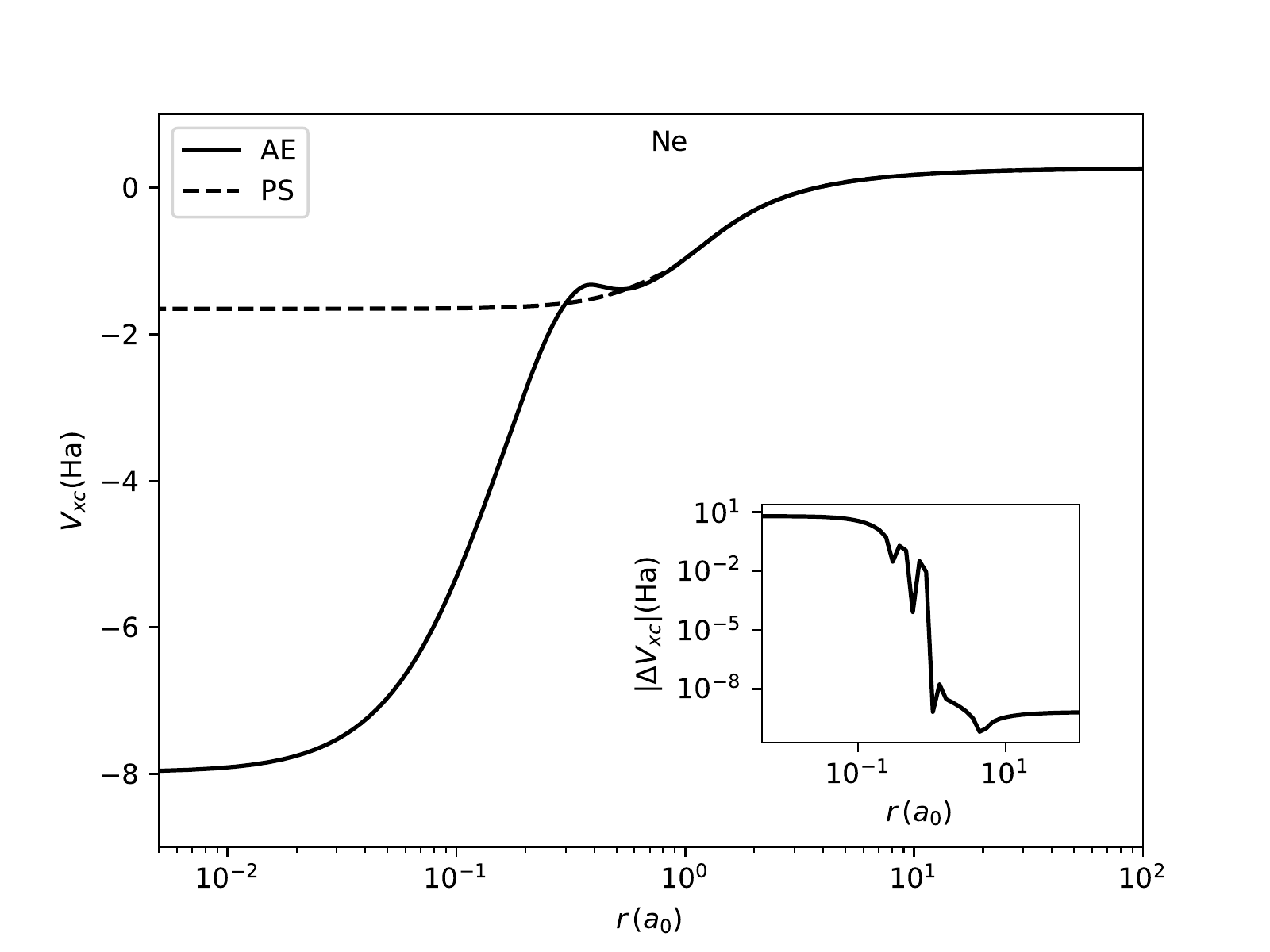} &
\includegraphics[width=0.8\columnwidth]{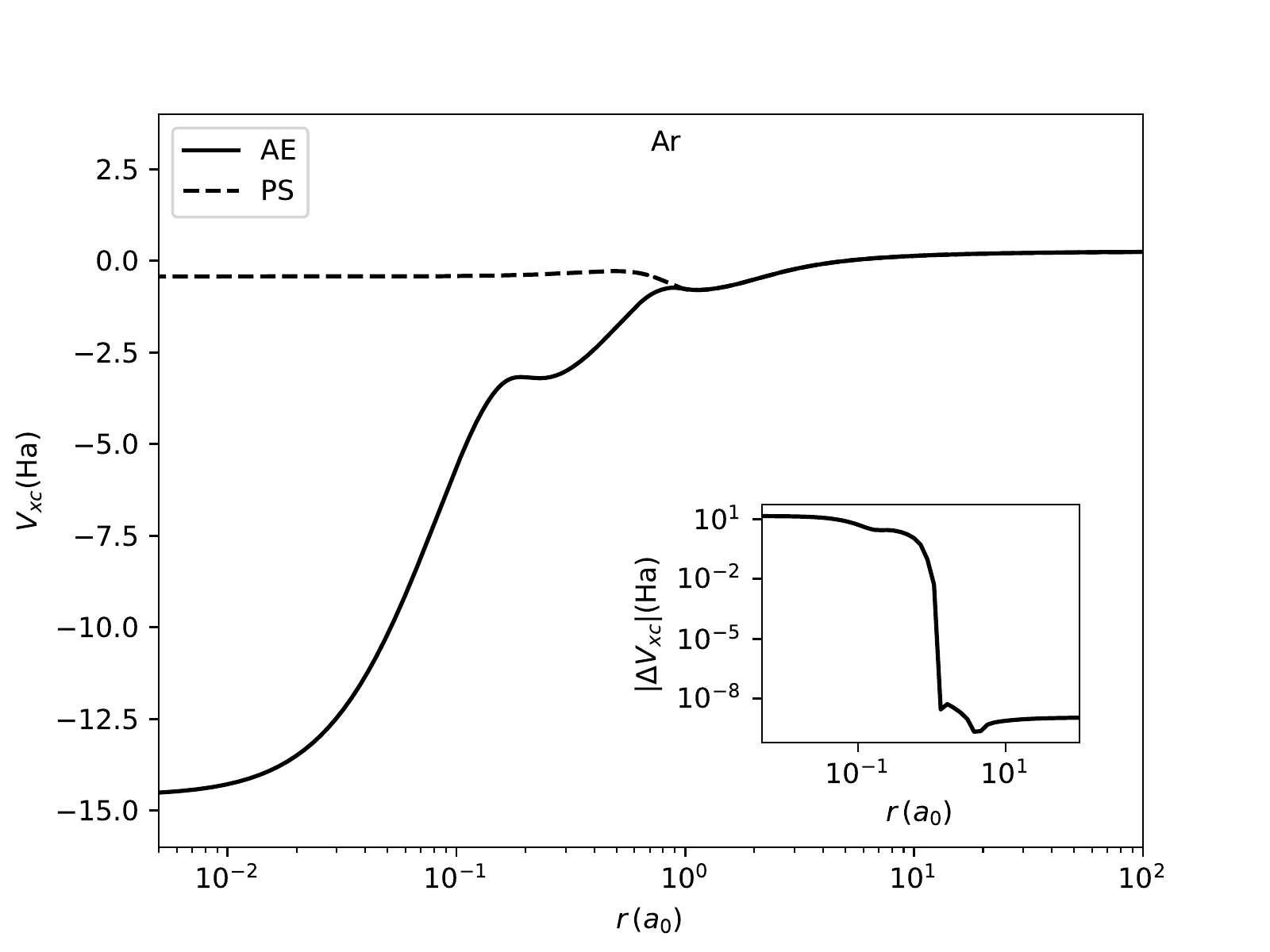} \\
\includegraphics[width=0.8\columnwidth]{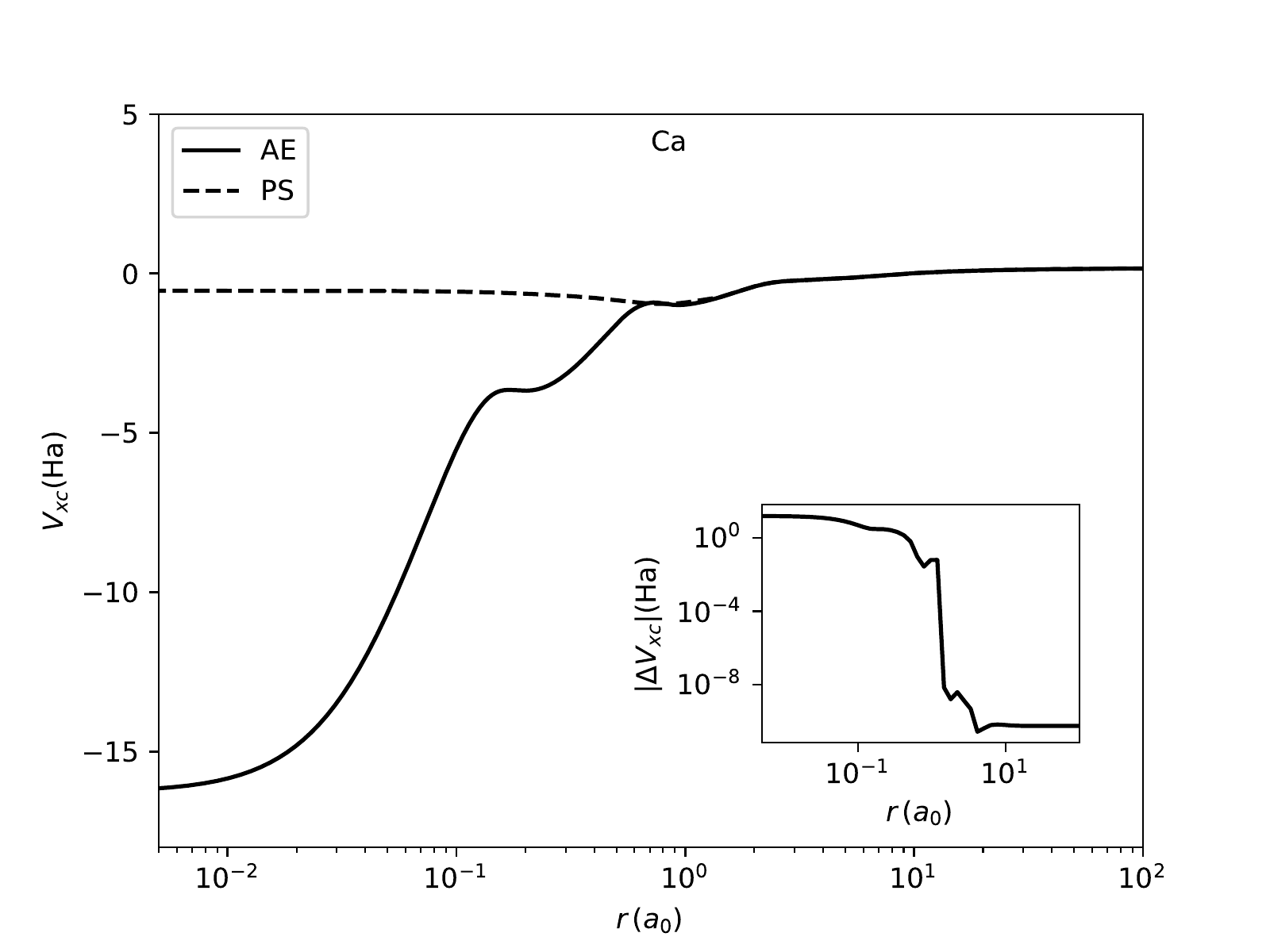} &
\includegraphics[width=0.8\columnwidth]{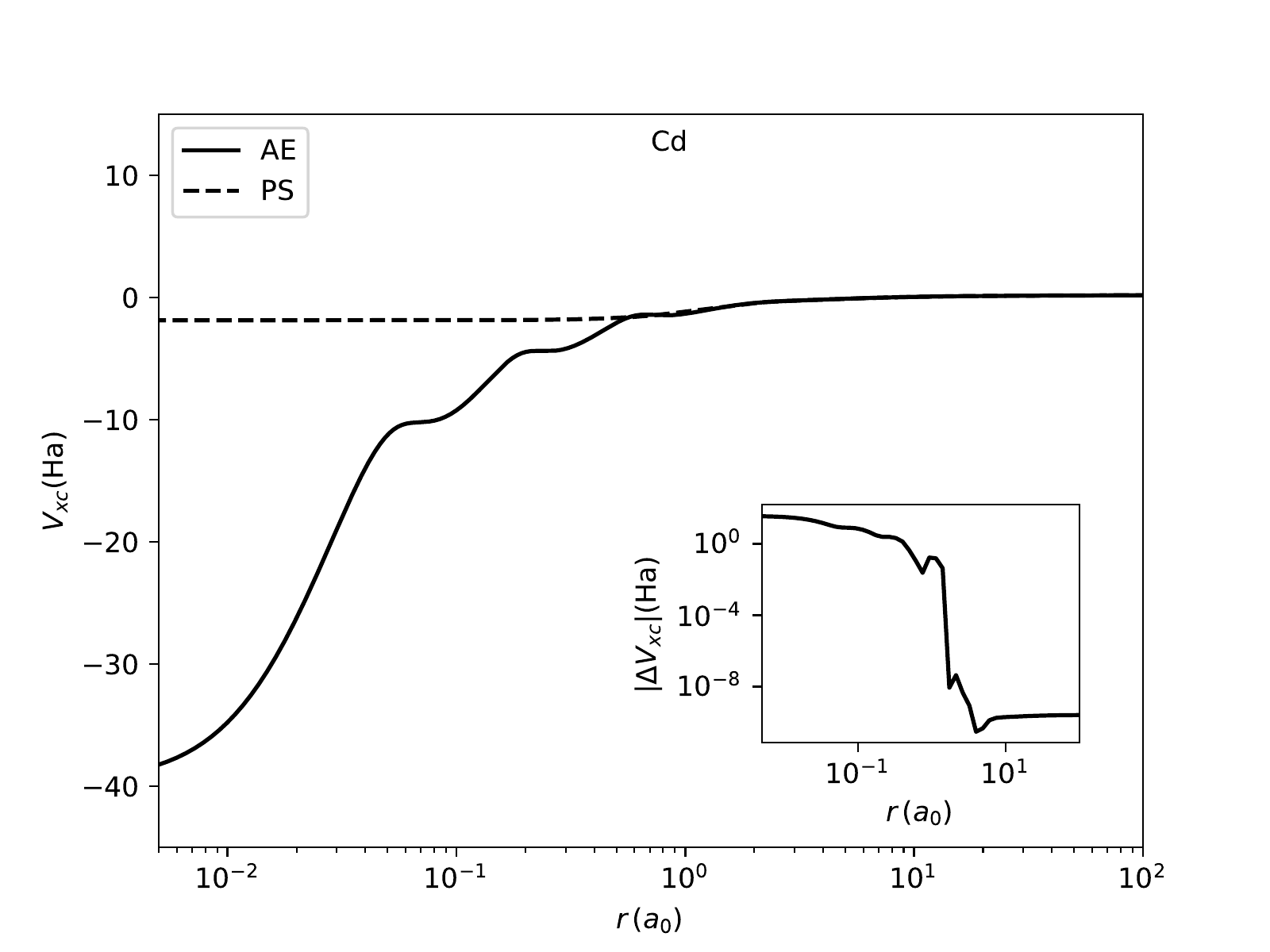}
\end{tabular}
\caption{The Becke-Johnson exchange and LDA correlation potential in a set of closed shell atoms, with all electrons included in the calculation (AE, solid line) and only valence electrons in a pseudopotential calculation (PS, dashed line). The inset shows the difference between the all-electron and pseudopotential calculation.}
\label{fig:atom_BJ_potentials}
\end{figure*}

Table~\ref{tab:singe_atom_energy} lists the orbital energies of a set of closed shell atoms, calculated by solving the KS equations for all electrons and for the valence electrons only, using ultrasoft pseudopotentials, with the Becke-Johnson exchange potential. The orbital energies of the valence states of the pseudo atoms show excellent agreement with their all-electron counterparts, within 2~meV across the range. We note that the energies of the semicore states included in Be, Ne and Mg show a larger deviation than the valence states, which is due to the fact that only a single ultrasoft projector was used for these states. However, the agreement is reasonable, and we expect that the discrepancy will not cause any significant effect in a practical calculation.

To demonstrate the accuracy of the pseudopotential, we compared the exchange-correlation potential functions from all-electron and pseudopotential calculations, on the same set of atoms. Figure~\ref{fig:atom_BJ_potentials} demonstrates that outside of $r_\textrm{c}$, the exchange-correlation potential curves match very accurately, and within $r_\textrm{c}$ the potential of the pseudo atom becomes a smooth function.

\subsection{Band structure calculations}

The main purpose of developing the Tran-Blaha exchange potential was to improve the description of the electronic band structure of solids within DFT, and in particular, the band gaps, which are often severely underestimated in LDA and GGA. An appropriate benchmark for the KED-including pseudopotential scheme is therefore comparing the calculated band structure to all-electron results. We used the selection of bulk semiconductor crystals with the GGA-optimized structural parameters from Ref.~\onlinecite{CamargoMartinez:2012kk}. We used the ELK software package \cite{elk}, a full-potential, linearized augmented plane-wave code, to perform the all-electron calculations, using the default species files. ELK employs a combination of local-orbital and augmented plane-wave basis functions, whose size was set by the \verb+vhighq+ keyword, which is recommended by the ELK manual to obtain highly converged results. The pseudopotential calculations were carried out with a modified version of \CASTEP{} 17.2. We used Monkhorst-Pack $k$-point grids\cite{Monkhorst:1976ta} with a 0.025~\AA$^{-1}$ spacing to sample the Brillouin zone, and the \verb+basis_precision : extreme+ setting in \CASTEP{} for the energy cutoff of the planewave basis.

The greatest shortcoming of pseudopotential calculations with the Tran-Blaha potential is that $\alpha$ and $\beta$ used in equation~\ref{eq:TB_c} are fitted based on the all-electron density, which is available neither at the point when the pseudopotentials are generated, nor during the calculation. In order to have access to the all-electron density and its gradient, they would need to be reconstructed, but this would not be practical on a Fourier grid representation. To benchmark our approach of generating and using pseudopotentials, we used the self-consistent values of $c$ obtained from the all-electron calculations. The parameter $c$ was fixed for both the pseudopotential generation and the electronic structure calculation. The results obtained using this approach are directly comparable to those of all-electron calculations, but for practical calculations, where $c$ is unknown, this method is clearly unfeasible.

To study the effect of employing different choices of $c$ at various stages of a calculation we performed band structure calculations with the following options, where $\tilde{c}$ is the value \emph{calculated} from the pseudo-density, and $c$ is the parameter used in calculating the exchange potential, as in equation~\ref{eq:MBJ}:
\begin{enumerate}[i]
    \item a series of runs where $c$ was set to $\tilde{c}$, calculated from the pseudo charge density, and allowed to vary self-consistently during the plane-wave calculation. Pseudopotentials were generated at $\tilde{c}_\textrm{SC}$ at the beginning of each run. This process is repeated until $\tilde{c}$ does not change between runs. This could be a realistic option in a production run for a compound where the all-electron $c$ is not available beforehand.
    \item Pseudopotentials generated using the all-electron $c_\textrm{AE}$, but $\tilde{c}$, as calculated from the pseudo charge density, was used as $c$, and allowed to vary self-consistently during the plane-wave calculation.
    \item Pseudopotentials generated using the Becke-Johnson potential, i.e. $c=1$, and $\tilde{c}$, as calculated from the pseudo charge density, was used as $c$ and allowed to vary self-consistently during the calculation.
    \item To demonstrate the effect of using pseudopotentials generated with a completely different class of functionals, we ran calculations with PBE pseudopotentials, using the appropriate all-electron $c_\textrm{AE}$ value in the plane-wave calculation.
\end{enumerate}
To give a general impression on how accurately we expect band gaps calculated from pseudopotential DFT match all-electron DFT, we also carried out calculations with the PBE functional, using the appropriate PBE pseudopotentials.

\begin{figure}[!ht]
    \includegraphics[width=\linewidth]{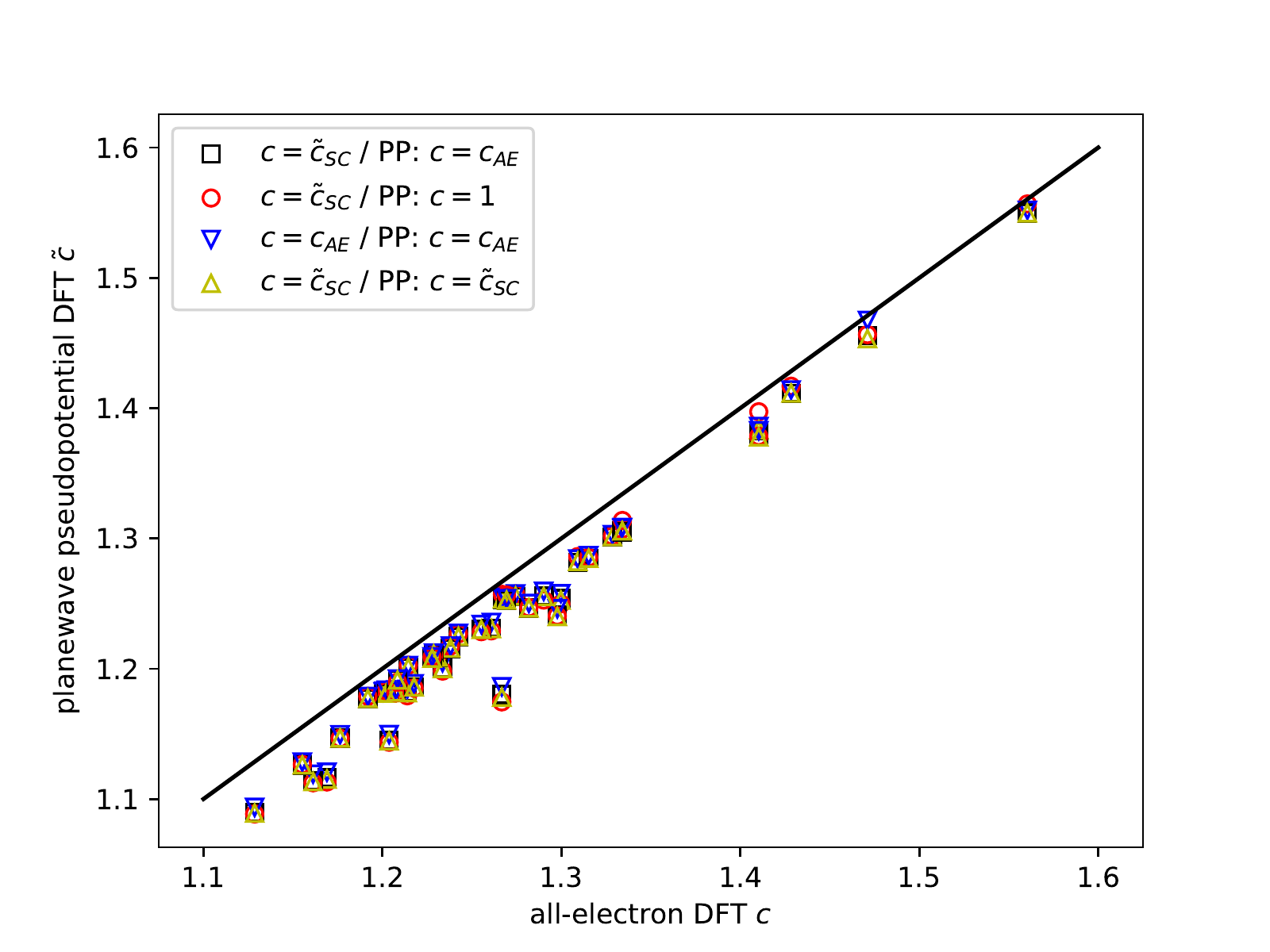}
    \caption{The Tran-Blaha $\tilde{c}$ parameter computed from the self-consistent pseudo charge density with ultrasoft pseudopotentials for all the semiconducting materials in our database, compared to the $c$ parameter obtained from all-electron calculations, using the Tran-Blaha exchange potential. Squares: the pseudopotentials were generated using the self-consistent $c$ parameter obtained from the all-electron calculations, circles: pseudopotentials generated with the BJ potential, downward triangles: $c$ fixed at the value obtained from all-electron calculation for both the pseudopotential generation and electronic structure calculation, upward triangles: final $c$ in a series of calculations for each compound where pseudopotentials for each run were generated using the self-consistent $\tilde{c}$.}
    \label{fig:TB_c}
\end{figure}

It is informative to examine the self-consistent values of $c$. Figure~\ref{fig:TB_c} compares the self-consistent $c$ of the pseudopotential calculations to those of the all-electron calculations. Even though there is a strong correlation, the pseudopotential $c$ values are consistently underestimated, with the Becke-Johnson type pseudopotentials being the furthest from the all-electron results. We note that it might be possible to refit the $\alpha$ and $\beta$ values in the expression for $c$ (equation~\ref{eq:TB_c}), but it is outside of the scope of this work, and such a reparameterization would be rather limited, being only applicable to a given set of pseudopotentials.

\begin{figure}[ht]
    \includegraphics[width=\linewidth]{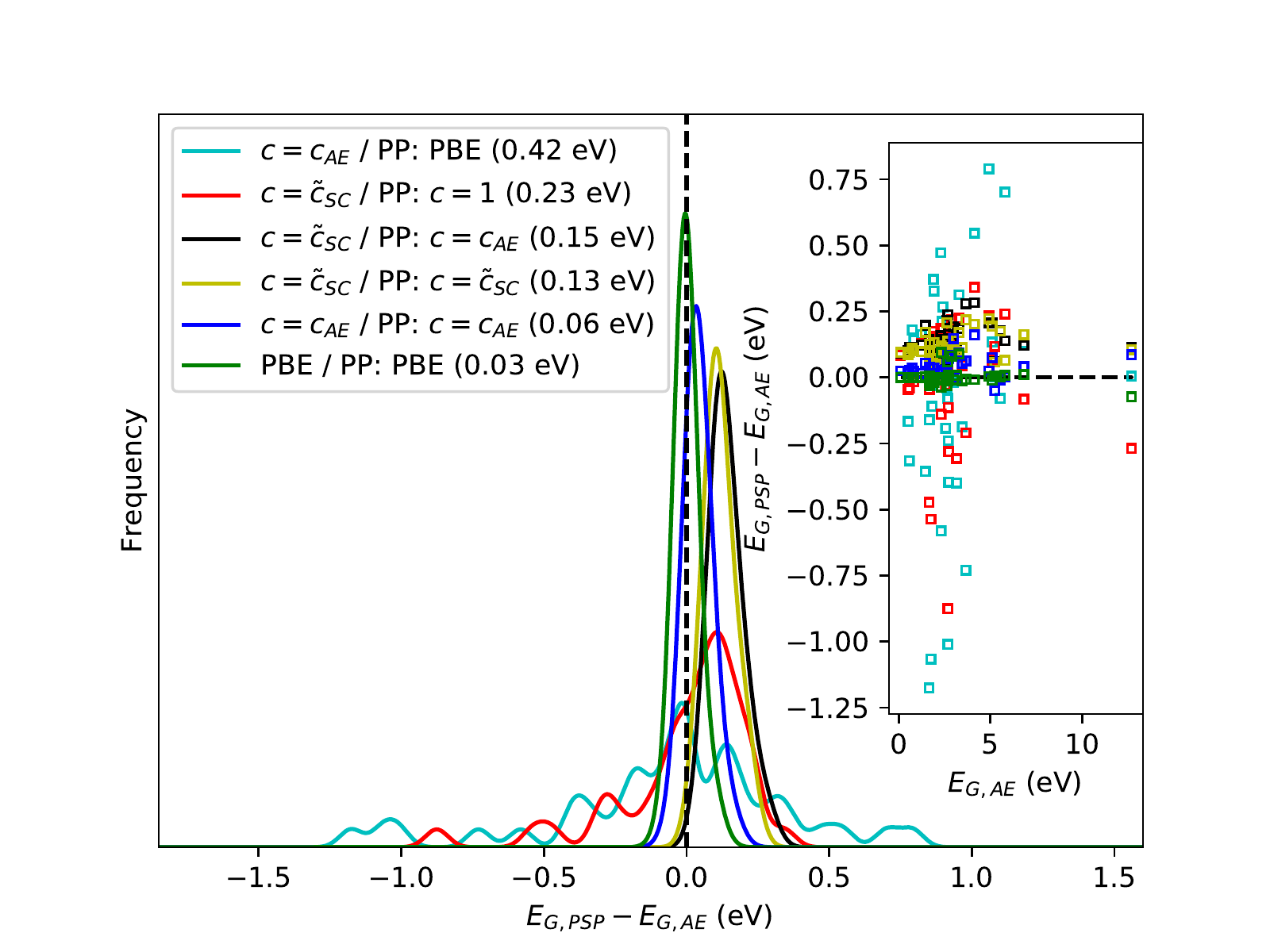}
    \caption{Distribution of errors in band gaps calculated with ultrasoft pseudopotentials ($E_\textrm{G,PSP}$), relative to all-electron calculations ($E_\textrm{G,AE}$). $c_\textrm{AE}$ indicates $c$ values fixed at the all-electron values, $\tilde{c}_\textrm{SC}$ means $c$ was allowed to vary self-consistently during the calculation, and we also included PBE reference calculations. PP defines the pseudopotential, based on the exchange potential used for generating it: PBE, Becke-Johnson ($c=1$), $c$ fixed at the all-electron values ($c=c_\textrm{AE})$ or the self-consistent $\tilde{c}_\textrm{SC}$. The numbers in brackets show the RMS error in the band gaps.\\
    The inset shows the errors of the calculations as function of the all-electron band-gaps.}
    \label{fig:band_gaps_PSP_LAPW}
\end{figure}

The calculated band gaps and self-consistent $c$ values are listed in Table~\ref{tab:band_gaps} and also shown in Figure~\ref{fig:band_gaps_PSP_LAPW}. With $c$ fixed throughout the calculation, we are able to reproduce the all-electron band gaps with our KED-enabled pseudopotentials with a root-mean-square error (RMSE) of 60 meV. For comparison, using the PBE GGA exchange-correlation functional, a similar performance is achieved at 30 meV RMSE. Being able to reproduce such a sensitive all-electron property is a good indication that the KED-supporting ultrasoft pseudopotentials are accurate and a viable alternative to all-electron and PAW calculations.

\begin{table*}[]
   \begin{threeparttable}[]
\begin{tabular}{lSS|SS|SS|SS|SS|SSS}
\hline\hline
{} & \multicolumn{2}{c}{AE(\mbj{})} &
\multicolumn{2}{c}{\makecell{$c = c_\textrm{AE}$ \\ PP: $c = c_\textrm{AE}$}} &
\multicolumn{2}{c}{\makecell{$c = \tilde{c}_\textrm{SC}$ \\ PP: $c = \tilde{c}_\textrm{SC}$}} &
\multicolumn{2}{c}{\makecell{$c = \tilde{c}_\textrm{SC}$ \\ PP: $c = c_\textrm{AE}$}} &
\multicolumn{2}{c}{\makecell{$c = \tilde{c}_\textrm{SC}$ \\ PP: $c = 1$}} &
\multicolumn{1}{c}{\makecell{$c = c_\textrm{AE}$ \\ PP: PBE}} & 
\multicolumn{1}{c}{AE(PBE)} & 
\multicolumn{1}{c}{\makecell{PBE \\ PP: PBE}} \\
\hline
{} &
   \multicolumn{1}{c}{$c$} &   \multicolumn{1}{c|}{$E_\textrm{G}$~(eV)} & 
   \multicolumn{1}{c}{$\tilde{c}$} &   \multicolumn{1}{c|}{$E_\textrm{G}$~(eV)} &
   \multicolumn{1}{c}{$\tilde{c}$} &   \multicolumn{1}{c|}{$E_\textrm{G}$~(eV)} &
   \multicolumn{1}{c}{$\tilde{c}$} &   \multicolumn{1}{c|}{$E_\textrm{G}$~(eV)} &
   \multicolumn{1}{c}{$\tilde{c}$} &   \multicolumn{1}{c|}{$E_\textrm{G}$~(eV)} &
   \multicolumn{1}{c}{$E_\textrm{G}$~(eV)} &
   \multicolumn{1}{c}{$E_\textrm{G}$~(eV)} &
   \multicolumn{1}{c}{$E_\textrm{G}$~(eV)} \\
\hline
InSb   & 1.20 &  0.10 &  1.18 &  0.07  &  1.18 &  0.00 &  1.18 &  0.00 &  1.18 &  0.01 &   0.09 &    0.00 &    0.00 \\
BN     & 1.30 &  5.79 &  1.26 &  5.79  &  1.25 &  5.73 &  1.25 &  5.65 &  1.25 &  \myred5.55 &   \myred5.09 &    4.46 &    4.45 \\
GaSb   & 1.20 &  0.58 &  1.18 &  0.56  &  1.18 &  0.49 &  1.18 &  0.47 &  1.18 &  0.62 &   \myred0.90 &    0.00 &    0.00 \\
MgO    & 1.43 &  6.83 &  1.41 &  6.78  &  1.41 &  6.66 &  1.41 &  6.70 &  1.42 &  6.91 &   6.70 &    4.44 &    4.43 \\
BaTe   & 1.20 &  2.31 &  1.18 &  2.22  &  1.18 &  2.20 &  1.18 &  2.18 &  1.18 &  2.21 &   2.20 &    1.66 &    1.57 \\
GaP    & 1.21 &  2.34 &  1.18 &  2.32  &  1.18 &  2.25 &  1.18 &  2.23 &  1.18 &  2.22 &   \myred2.13 &    1.56 &    1.57 \\
   MgS\tnote{a}    & 1.23 &  4.13 &  1.20 &  3.97  &  1.20 &  \myred3.93 &  1.20 &  \myred3.85 &  1.20 &  \myred3.79 &   \myred3.59 &    2.74 &    2.75 \\
   GaN\tnote{b}    & 1.33 &  2.71 &  1.30 &  2.64  &  1.30 &  2.61 &  1.30 &  2.54 &  1.30 &  \myred2.99 &   \myred3.11 &    1.49 &    1.49 \\
AlAs   & 1.18 &  2.23 &  1.15 &  2.21  &  1.15 &  2.18 &  1.15 &  2.14 &  1.15 &  2.14 &   2.23 &    1.51 &    1.51 \\
BP     & 1.17 &  1.89 &  1.12 &  1.89  &  1.12 &  1.79 &  1.12 &  1.77 &  1.11 &  1.74 &   \myred1.52 &    1.26 &    1.27 \\
MgSe   & 1.21 &  2.99 &  1.20 &  2.84  &  1.20 &  2.88 &  1.20 &  2.80 &  1.20 &  2.86 &   3.01 &    1.76 &    1.77 \\
CdSe   & 1.27 &  1.92 &  1.26 &  1.88  &  1.26 &  1.79 &  1.26 &  1.80 &  1.26 &  1.74 &   \myred1.59 &    0.49 &    0.50 \\
SiC    & 1.20 &  2.29 &  1.15 &  2.27  &  1.14 &  2.22 &  1.14 &  2.13 &  1.14 &  2.11 &   \myred1.82 &    1.38 &    1.38 \\
LiF    & 1.56 & 12.69 &  1.55 & 12.60  &  1.55 & 12.58 &  1.55 & 12.57 &  1.56 & \myred12.96 &  12.68 &    8.75 &    8.83 \\
BAs    & 1.21 &  1.75 &  1.18 &  1.73  &  1.18 &  1.71 &  1.18 &  1.68 &  1.18 &  1.71 &   1.78 &    1.21 &    1.22 \\
BaS    & 1.26 &  3.29 &  1.24 &  3.19  &  1.23 &  3.12 &  1.23 &  3.10 &  1.23 &  \myred3.06 &   \myred2.97 &    2.24 &    2.15 \\
ZnTe   & 1.23 &  2.31 &  1.21 &  2.28  &  1.21 &  2.19 &  1.21 &  2.17 &  1.21 &  2.45 &   \myred2.89 &    1.05 &    1.09 \\
AlN    & 1.30 &  5.53 &  1.25 &  5.54  &  1.24 &  5.36 &  1.24 &  5.36 &  1.24 &  5.53 &   5.61 &    4.17 &    4.17 \\
C      & 1.27 &  4.92 &  1.19 &  4.89  &  1.18 &  \myred4.69 &  1.18 &  \myred4.71 &  1.17 &  \myred4.69 &   \myred4.13 &    4.13 &    4.12 \\
MgS\tnote{b} & 1.26 &  5.09 &  1.23 &  5.01  &  1.23 &  4.89 &  1.23 &  \myred4.88 &  1.23 &  5.02 &   4.95 &    3.35 &    3.36 \\
InP    & 1.22 &  1.45 &  1.19 &  1.40  &  1.19 &  1.28 &  1.19 &  1.25 &  1.19 &  1.40 &   \myred1.81 &    0.45 &    0.45 \\
GaN\tnote{c} & 1.33 &  3.15 &  1.31 &  3.15  &  1.31 &  3.05 &  1.31 &  3.04 &  1.31 &  \myred3.46 &   \myred3.55 &    1.91 &    1.91 \\
CuCl   & 1.33 &  1.76 &  1.31 &  1.77  &  1.31 &  1.70 &  1.30 &  1.66 &  1.31 &  \myred2.29 &   \myred2.82 &    0.51 &    0.52 \\
Si     & 1.13 &  1.24 &  1.09 &  1.23  &  1.09 &  1.14 &  1.09 &  1.12 &  1.09 &  1.11 &   1.08 &    0.62 &    0.62 \\
BaSe   & 1.24 &  2.89 &  1.23 &  2.81  &  1.22 &  2.78 &  1.22 &  2.76 &  1.23 &  \myred2.67 &   2.81 &    2.03 &    1.95 \\
CuBr   & 1.31 &  1.65 &  1.28 &  1.61  &  1.28 &  1.59 &  1.28 &  1.51 &  1.29 &  \myred2.12 &   \myred2.83 &    0.41 &    0.43 \\
CdS    & 1.29 &  2.63 &  1.26 &  2.58  &  1.26 &  \myred2.43 &  1.26 &  \myred2.42 &  1.25 &  2.48 &   2.67 &    1.03 &    1.06 \\
CdTe   & 1.24 &  1.69 &  1.22 &  1.66  &  1.22 &  1.57 &  1.22 &  1.57 &  1.22 &  1.59 &   1.70 &    0.58 &    0.61 \\
MgTe   & 1.19 &  3.46 &  1.18 &  3.41  &  1.18 &  3.35 &  1.18 &  3.35 &  1.18 &  3.42 &   3.65 &    2.30 &    2.31 \\
AlP    & 1.16 &  2.40 &  1.12 &  2.39  &  1.11 &  2.29 &  1.11 &  2.26 &  1.11 &  2.22 &   \myred2.14 &    1.63 &    1.64 \\
GaAs   & 1.23 &  1.67 &  1.21 &  1.64  &  1.21 &  1.57 &  1.21 &  1.55 &  1.21 &  1.72 &   1.83 &    0.52 &    0.52 \\
AgI    & 1.27 &  2.67 &  1.25 &  2.59  &  1.25 &  2.56 &  1.25 &  2.54 &  1.26 &  2.63 &   2.75 &    0.90 &    0.91 \\
AgF    & 1.47 &  2.55 &  1.47 &  2.54  &  1.45 &  2.45 &  1.46 &  2.49 &  1.46 &  2.58 &   2.75 &   -0.34 &   -0.34 \\
ZnO    & 1.41 &  2.67 &  1.39 &  2.59  &  1.38 &  2.53 &  1.38 &  \myred2.43 &  1.40 &  \myred3.55 &   \myred3.68 &    0.86 &    0.80 \\
InN    & 1.32 &  0.82 &  1.29 &  0.80  &  1.28 &  0.72 &  1.28 &  0.72 &  1.29 &  0.84 &   0.68 &    0.02 &    0.02 \\
ZnS    & 1.28 &  3.66 &  1.25 &  3.60  &  1.25 &  \myred3.44 &  1.25 &  \myred3.38 &  1.25 &  \myred3.87 &   \myred4.39 &    1.99 &    2.00 \\
InAs   & 1.23 &  0.74 &  1.21 &  0.70  &  1.21 &  0.62 &  1.21 &  0.62 &  1.21 &  0.62 &   0.56 &    0.00 &    0.00 \\
Ge     & 1.21 &  0.51 &  1.19 &  0.50  &  1.19 &  0.43 &  1.19 &  0.41 &  1.19 &  0.56 &   0.68 &    0.00 &    0.00 \\
CaO    & 1.41 &  5.24 &  1.38 &  5.29  &  1.38 &  5.18 &  1.38 &  5.18 &  1.38 &  5.12 &   5.24 &    3.66 &    3.66 \\
ZnSe   & 1.27 &  2.70 &  1.25 &  2.63  &  1.25 &  2.57 &  1.25 &  2.54 &  1.26 &  2.82 &   \myred2.94 &    1.13 &    1.13 \\
AlSb   & 1.16 &  1.81 &  1.13 &  1.80  &  1.13 &  1.76 &  1.13 &  1.73 &  1.13 &  1.77 &   1.92 &    1.24 &    1.24 \\
\hline
RMSE  &       &       &       &  0.06  &      &   0.13 &       & 0.15 &        &  0.23 &   0.42 &         & 0.03 \\
\bottomrule
\end{tabular}
    \caption{Calculated band gaps and self-consistent $c$ values of a selection
    of semiconductors. AE(\mbj{}) and AE(PBE) are all-electron calculations using
    the Tran-Blaha potential and PBE, respectively. Pseudopotential generation
    and electronic structure calculations were run at $c$ fixed at the
    all-electron self-consistent value $c_\textrm{AE}$, or 1 (corresponding to
    the Becke-Johnson potential) or $c$ was let to vary during the calculation
    ($\tilde{c}_\textrm{SC}$).\tnote{d} We also present results where the PBE pseudopotentials
    were used. Highlighted are band gap values where the difference between the all-electron and pseudopotential calculation was greater than 0.2~eV.}
    \label{tab:band_gaps}
    \begin{tablenotes}
    \item [a] rocksalt structure
    \item [b] zincblende structure
    \item [c] wurtzite structure
    \item [d] Note that in the first two cases the self-consistent $\tilde{c}$, an output of the calculation, does not necessarily coincide with the fixed $c$ value, used in the calculation of the exchange potentials.
    \end{tablenotes}
   \end{threeparttable}
\end{table*}

In practice, it is not uncommon to use pseudopotentials in a calculation which were generated with mismatching exchange-correlation functionals. We studied the effect of the choice of exchange-correlation functional in the pseudopotential by calculating the band gaps on the same set of materials, using pseudopotentials generated with PBE, and with $c$ fixed at the corresponding all-electron value. It is striking that this approach produces an order of magnitude less accurate results on average, with some significant outliers.

In the general case, as we mentioned above, the all-electron $c$ is not available in advance. Therefore we studied the more realistic scenario when $c$, computed from the pseudo-density, is allowed to vary self-consistently during the calculations. This leads to less accurate band gaps compared to the all-electron results, due to the fact that $c$ no longer matches the all-electron $c$; effectively a different exchange potential is being used. In actuality, using pseudopotentials generated with the Becke-Johnson potential leads to a loss of accuracy, but even self-consistent calculations can be improved significantly if pseudopotentials are generated at $c$ set to the corresponding all-electron value.

In order to study a material where $c$ is unknown, for practical purposes, one could employ a procedure when calculations are `bootstrapped' using pseudopotentials generated with the Becke-Johnson exchange potential, and iteratively improving $c$ by each time reconstructing the pseudopotential with the new $c$, until self-consistence is achieved. According to our tests, self-consistence is achieved in a few steps, and being able to reuse the densities from the previous calculation means the increase of the cost of computation is not significant. The accuracy of this approach is comparable to that of using pseudopotentials generated at $c_\textrm{AE}$, making it a viable option for practical calculations.

\begin{figure}
    \includegraphics[width=\linewidth]{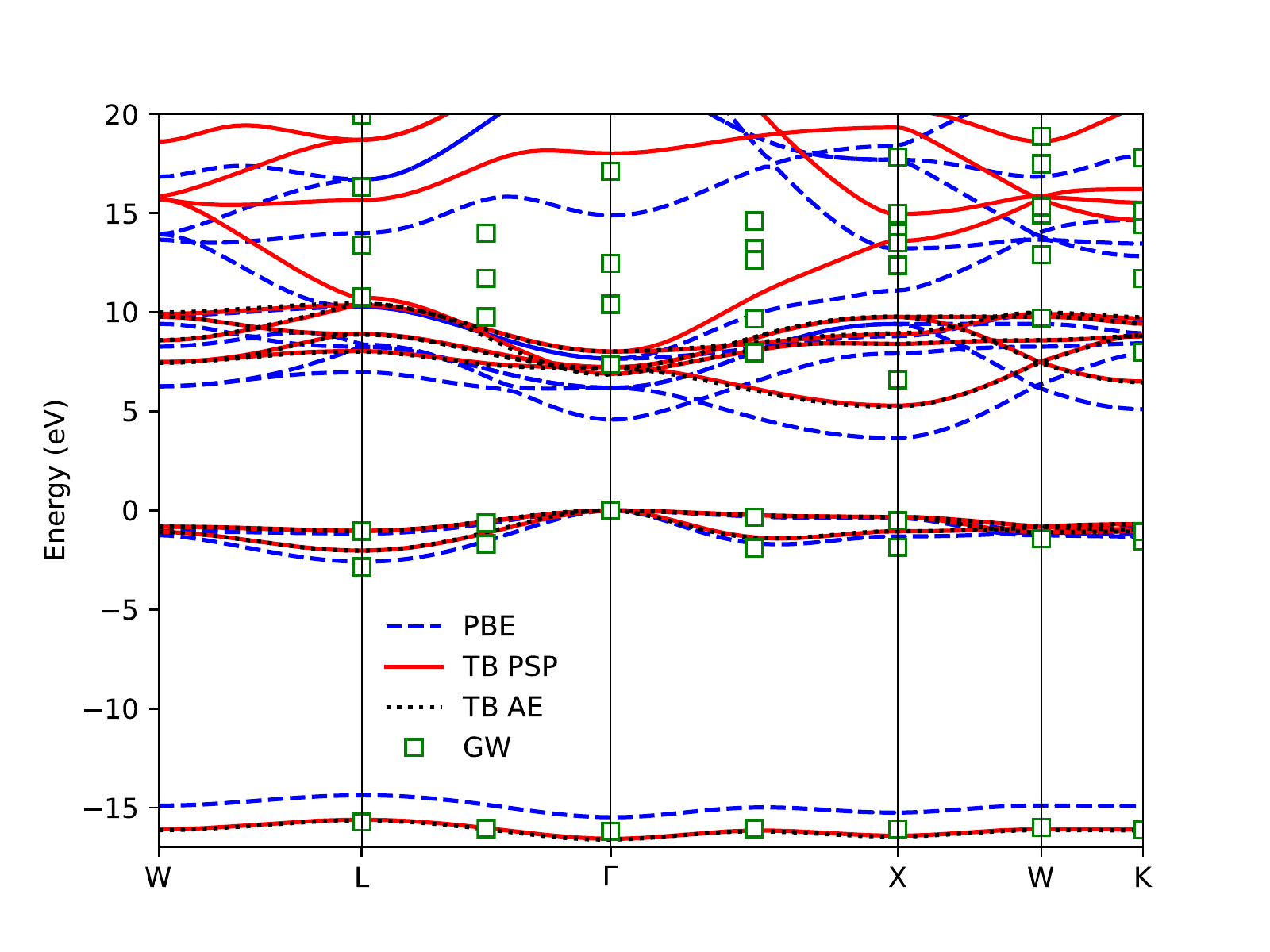}
    \caption{The band structure of CaO along high-symmetry lines. Blue dashed lines calculated by PBE,  red solid lines are the result of the Tran-Blaha exchange potential, with $c=1.410$ (the all-electron value), black dotted lines are the all-electron results with the Tran-Blaha exchange, green squares represent the GW values\cite{Yamasaki:2002bk}.}
    \label{fig:CaO_bandstructure}
\end{figure}

We also studied how well the dispersion of the band energies can be reproduced in the pseudopotential calculation. As an example, the band structure of CaO is plotted on Figure~\ref{fig:CaO_bandstructure}, with a the all-electron and pseudopotential calculations matching to the width of the line. For comparison, we added the PBE and GW\cite{Yamasaki:2002bk} results, to illustrate the vast improvement in the location of the conduction bands, but also to show that the shape of the conduction bands is somewhat less well reproduced, this being a general feature of the \mbj{} potential, not a result of the pseudopotential approximation.

\subsection{NMR calculations}

NMR is an often used experimental technique to determine the atomistic structure of matter. Assisting this, NMR parameters are routinely calculated from first principles, and DFT calculations have been found to be reliable for a wide range of systems. In solids, DFT calculations with GGA functionals, combined with GIPAW\cite{Pickard:2001eua,Yates:2007ic} show remarkable accuracy for isotropic shielding and J-coupling parameters, with a few notable exceptions, for example, fluorides.

\begin{figure}
    \includegraphics[width=\linewidth]{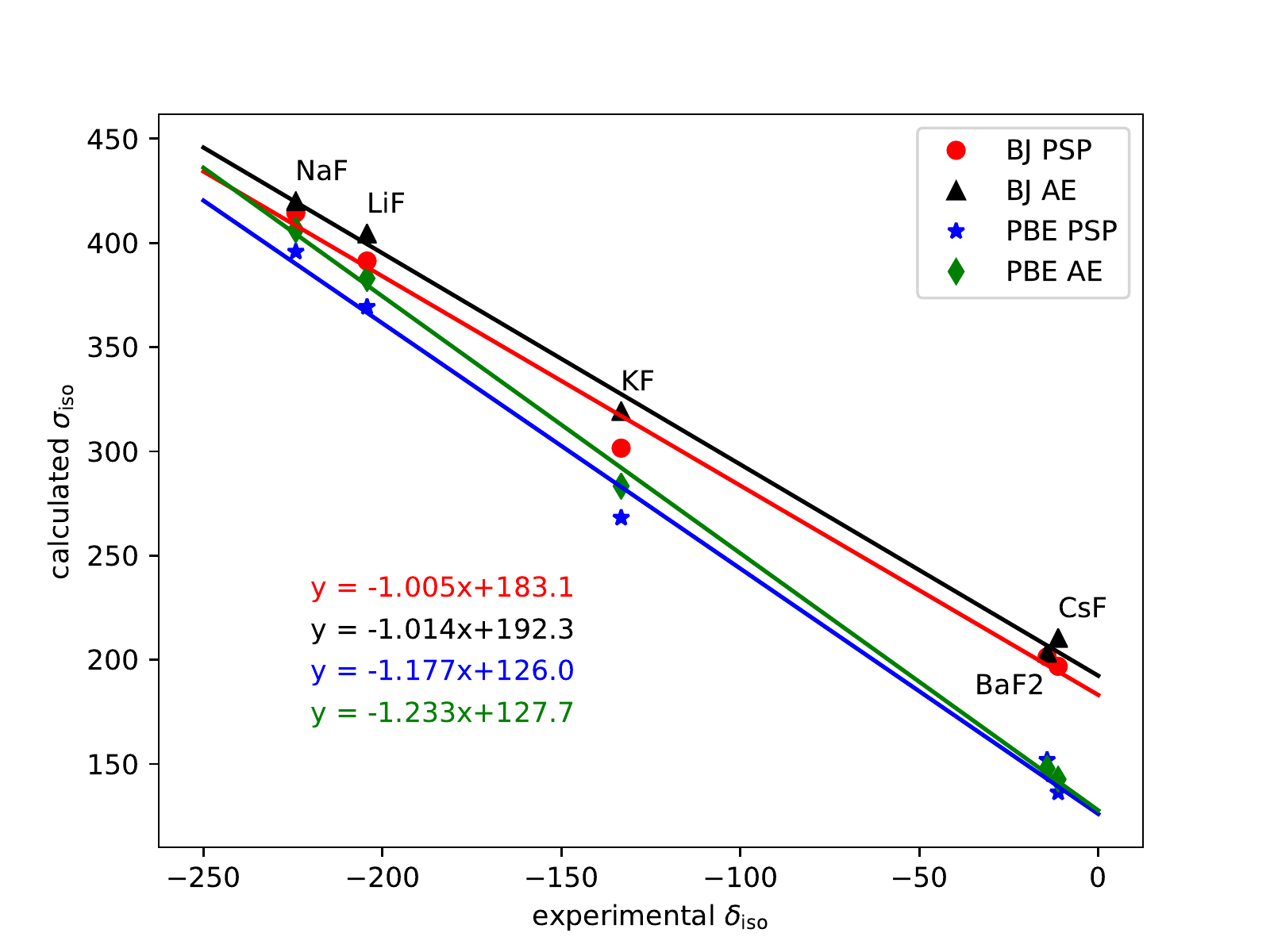}
    \caption{Calculated isotropic shieldings of $^{19}\textrm{F}$ compared to experimental isotropic shifts. Red circles and black triangles are results obtained with the Becke-Johnson exchange potential, in pseudopotential and all-electron calculation\cite{Laskowski:2013jm}, respectively. Blue stars and green diamonds are PBE results obtained with pseudopotential GIPAW\cite{Sadoc:2011cl} and all-electron calculations\cite{Laskowski:2013jm}, respectively. }
    \label{fig:NMR_fluorides}
\end{figure}

The Becke-Johnson exchange potential was shown to improve the NMR shielding in a
set of inorganic fluoride compounds\cite{Laskowski:2013jm}, using an
all-electron approach to solve the electronic structure problem. In this current
work, we use some of these results to validate our implementation of
KED-supporting pseudopotentials. We computed the $^{19}\textrm{F}$ NMR
shieldings of LiF, NaF, KF, CsF and $\textrm{BaF}_2$ with the Becke-Johnson
exchange potential using the GIPAW method as implemented in \CASTEP{}. The results are summarized in Figure~\ref{fig:NMR_fluorides}, showing that we are able to reproduce the slope of the shielding vs. experimental shift points of the all-electron calculations.

\section{Conclusion}
Meta-generalized gradient approximation functionals are gaining popularity in electronic structure calculations, with some implementations providing considerable improvements over GGA functionals\cite{Sun:2015ef}. An often used input variable is the kinetic energy density, which needs to be pseudized in a plane-wave basis to enable practical calculations. We present a scheme which extends ultrasoft pseudopotentials to support kinetic energy densities, and implemented it in the \CASTEP{} code. We have carried out calculations to benchmark the performance and reliability of KED-enabled pseudopotentials in a range of systems, showing that all-electron results can be reproduced accurately, given the exchange potential can be kept consistent. We note that this mechanism of generating pseudopotentials can be extended in a straightforward manner to mGGA functionals, as demonstrated by our related work on the SCAN functional\cite{bartok2019scan}.

More specifically, regarding the Tran-Blaha potential, we found that the $c$ parameter, responsible to inform the potential about the global electronic structure, cannot be reproduced precisely using the pseudo-density. This somewhat limits the usability of the Tran-Blaha potential in a pseudopotential calculation, as the exchange potential becomes implicitly pseudopotential dependent. We suggest practical workarounds this problem, and provided extensive benchmarks to inform the community on the expected accuracy and reliability of these options. Another avenue, although outside of the scope of this work, would be reparameterization of the Tran-Blaha potential, based on pseudo densities.

Finally, we explored the work flow of using pseudopotentials generated with different exchange-correlation functional or potential than the the one used in the self-consistent calculation. We found that this approach leads to significant differences in the resulting electronic structure compared to all-electron calculations, and conclude that this practice should be avoided.

\begin{acknowledgments}
The authors would like to thank Chris Pickard, Philip Hasnip and Dominik Jochym for useful discussions, and Ewan Richardson for helpful suggestions. ABP acknowledges support from the Collaborative Computational Project for NMR Crystallography (CCP-NC) and UKCP Consortium, both funded by the Engineering and Physical Sciences Research Council (EPSRC) under grant numbers EP/M022501/1 and EP/P022561/1, respectively. Computing resources were provided by the STFC Scientific Computing Department's SCARF cluster.
\end{acknowledgments}

\bibliography{TB}

\end{document}